\documentclass[preprint, modern]{aastex631}

\usepackage{amsmath, amssymb, amsfonts}
\usepackage{mathtools}
\usepackage{xspace}
\usepackage{multirow}

\newcommand{\Msun}{\ensuremath{\mathrm{M}_\odot}\xspace}
\newcommand{\Rsun}{\ensuremath{\mathrm{R}_\odot}\xspace}
\newcommand{\br}{\mathrm{br}}
\newcommand{\ej}{\mathrm{ej}}
\newcommand{\sh}{\mathrm{sh}}
\newcommand{\tr}{\mathrm{tr}}

\newcommand{\diff}{\mathrm{diff}}
\newcommand{\crit}{\mathrm{crit}}
\newcommand{\tem}{t_\mathrm{em}}
\newcommand{\Lint}{L_\mathrm{int}}
\newcommand{\CSM}{\mathrm{CSM}}
\newcommand{\odv}[2]{\frac{\mathrm{d} #1}{\mathrm{d} #2}}
\newcommand{\odif}[1]{\mathrm{d} #1}

\submitjournal{ApJ}

\shorttitle{Interacting SNe with flat CSM}
\shortauthors{Chiba et al.}

\graphicspath{{./}{/}}

\begin{document}

\title{Characterisation of Supernovae Interacting with Dense Circumstellar Matter with a Flat Density Profile}

\correspondingauthor{Ryotaro Chiba}
\email{ryotaro.chiba@grad.nao.ac.jp}

\author[0009-0003-4594-3715]{Ryotaro Chiba}
\affiliation{Astronomical Science Program, Graduate Institute for Advanced Studies, SOKENDAI \\
2-21-1 Osawa, Mitaka, Tokyo 181-8588, Japan}
\affiliation{National Astronomical Observatory of Japan, National Institutes of Natural Sciences \\
2-21-1 Osawa, Mitaka, Tokyo 181-8588, Japan}

\author[0000-0003-1169-1954]{Takashi J. Moriya}
\affiliation{National Astronomical Observatory of Japan, National Institutes of Natural Sciences \\
2-21-1 Osawa, Mitaka, Tokyo 181-8588, Japan}
\affiliation{Astronomical Science Program, Graduate Institute for Advanced Studies, SOKENDAI \\
2-21-1 Osawa, Mitaka, Tokyo 181-8588, Japan}
\affiliation{School of Physics and Astronomy, Monash University, Clayton, Victoria 3800, Australia}

\begin{abstract}
Interaction between supernova (SN) ejecta and dense circumstellar medium (CSM) with a flat density structure ($\rho \propto r^{-s}, s < 1.5$) was recently proposed as a possible mechanism behind interacting SNe that exhibit exceptionally long rise times exceeding 100 days.
In such a configuration, the interaction luminosity keeps rising until the reverse shock propagates into the inner layers of the SN ejecta.
We investigate the light curves of SNe interacting with a flatly distributed CSM in detail, incorporating the effects of photon diffusion inside the CSM into the model.
We show that three physical processes -- the shock breakout, the propagation of the reverse shock into the inner ejecta, and the departure of the shock from the dense CSM -- predominantly determine the qualitative behaviour of the light curves.
Based on the presence and precedence of these processes, the light curves of SNe interacting with flatly distributed CSM can be classified into five distinct morphological classes.
We also show that our model can qualitatively reproduce doubly peaked SNe whose peaks are a few tens of days apart, such as SN 2005bf and SN 2022xxf.
Our results show that the density distribution of the CSM is an important property of CSM that contributes to the diversity in light curves of interacting SNe.
\end{abstract}


\section{Introduction} \label{sec:intro}

It is estimated that around $10 \%$ of core-collapse (CC) supernovae (SNe) are classified as ``interacting'' SNe; those of Type IIn, Ibn or Icn \citep[e.g.][]{Perley2020-gc} that display signatures of interaction between the SN ejecta and a dense circumstellar matter (CSM; see \citealt{Fraser2020-xt} for a recent review).
The letter ``n'' in these classifications indicates the presence of narrow line components in their spectra \citep{Filippenko1997-sj} that arise from the interaction of the SN ejecta with the dense CSM \citep[e.g.][]{Chugai1994-us, Chugai2001-qy, Chugai2004-xx}.

Light curves of interacting SNe exhibit vast diversity \citep[e.g.][]{Kiewe2012-xk, Taddia2013-bs, Nyholm2020-ab}, resulting from differences in various CSM properties, such as its mass, location, and density profile \citep[e.g.][]{Villar2017-ux, Suzuki2020-wl}.
For example, some interacting SNe are much brighter than non-interacting SNe, and the luminosity enhancement can be attributed to efficient dissipation of ejecta kinetic energy via interaction between ejecta and dense CSM.
Furthermore, many interacting SNe rise to their peak luminosity in the timescale of $\sim 50$ days \citep[][]{Nyholm2020-ab}, much slower than usual non-interacting SNe.
Typically, this prolonged evolution is explained as the diffusion timescale of photons inside optically thick CSM \citep[e.g.][]{Smith2007-pw, Chevalier2011-xw, Moriya2013-lj}.
Meanwhile, some interacting SNe display extremely long rise times of a few hundred days \citep[e.g.][]{Miller2010-ju, Smith2024-ul}.
If these long rise times were accomplished as diffusion time in a CSM, the required mass of the CSM is $\gtrsim$ 100 \Msun \citep[e.g.][]{Suzuki2020-wl, Khatami2023-oy}.
Such a massive CSM is difficult to form.

In most of the previous studies, the interaction between the SN ejecta and the CSM was investigated under the assumption of a wind-like density profile of the CSM ($\rho_\mathrm{CSM} \propto r^{-2}$).
However, stellar wind is not the only channel of CSM formation; depending on the formation process, the CSM density profile may vary.
So far, barring some exceptions \citep[e.g.][]{Moriya2012-tu,Chatzopoulos2012-ig}, the effect of diversity in CSM density structures on the interacting SNe has yet to be investigated much.
Recently, \citet{Moriya2023-kk} suggested that interacting SNe with light curves having long rise times of a few hundred days might be related to a flatter CSM density profile ($\rho_\mathrm{CSM} \propto r^{-s}$ where $s < 1.5$). 
Such a density profile allows the luminosity from the ejecta--CSM interaction to increase over time, as long as the CSM interacts with the outer layers of the ejecta.

In this work, we extend the model of \citet{Moriya2023-kk}, which was discussed based on the following assumptions:
First, the optical depth of the unshocked CSM medium was assumed to be low enough so that photon diffusion inside the CSM does not affect the light curves.
Second, it was assumed that the CSM extends far enough so that the inner ejecta begins interacting with the CSM before the ejecta reaches the edge of the CSM.
We investigate how the landscape of the light curves from interacting SNe inside a flatly distributed CSM changes when the effects of diffusion and the extent of the CSM are considered.

This paper is structured as follows.
In Section~\ref{sec:method}, we present our model for describing interacting SNe that account for the effects of photon diffusion.
We also outline the physical processes during the ejecta--CSM interaction that determine the shapes of the light curves.
In Section~\ref{sec:morphology}, we introduce the classification scheme of the light curves in our model based on their qualitative behaviours.
We also calculate how the timescales and luminosities of the light curves change with respect to the mass and the size of the CSM in our model with flat CSM.
We then demonstrate that a flat CSM can result in doubly peaked light curves of interacting SNe and compare our model with some doubly peaked SN light curves in the literature.
Finally, in Section~\ref{sec:summary}, we summarise our findings.

\section{Model Description} \label{sec:method}

We consider SNe whose dominant source of luminosity is the interaction between SN ejecta and dense CSM.
We assume spherical symmetry of the system and consider a single, CSM shell of mass $M_\CSM$ and radius $R_\CSM$ that begins to interact with the ejecta at $t = 0$.

In Section \ref{sec:evolution}, we outline the general formalism adopted to determine the evolution of interacting SNe as in previous studies \citep[e.g.][]{Chatzopoulos2012-ig, Moriya2013-sp, Hiramatsu2024-hb}.
In Section \ref{sec:diffusion}, we describe how the effects of diffusion are treated in our model.

\subsection{Evolution of Shocked Shell} \label{sec:evolution}

\begin{figure}[ht!]
    \centering
    \includegraphics[width = 0.70 \textwidth]{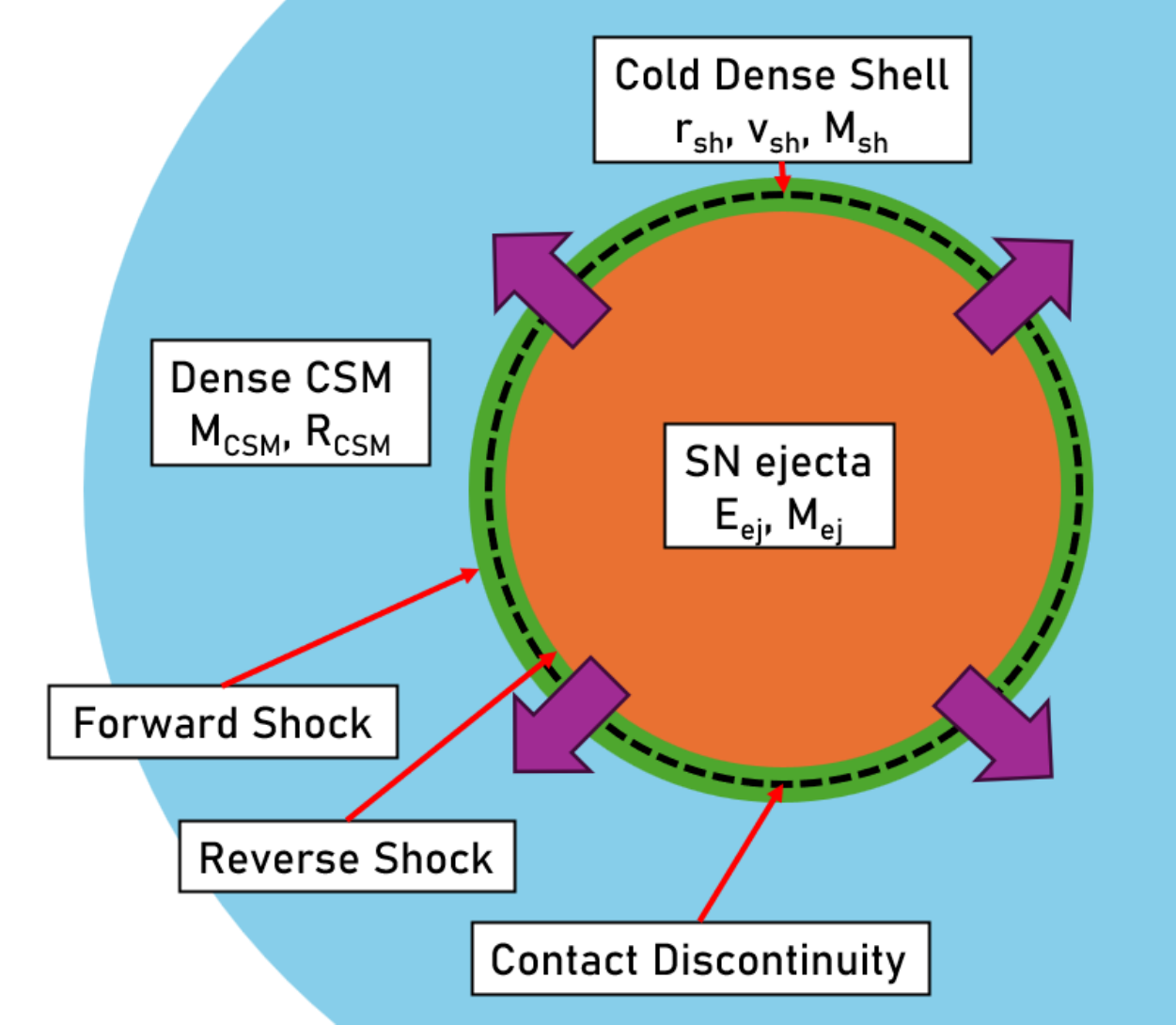}
    \caption{Schematic illustration of the structure of the radiative shock formed by the collision between the ejecta and CSM. The cold dense shell is located at a radius of $r_\sh$.}
    \label{fig:shock_structure}
\end{figure}

The collision between the SN ejecta and the CSM launches two shock waves from the contact discontinuity: a forward shock that propagates into the CSM and a reverse shock that propagates into the ejecta, as illustrated in Figure \ref{fig:shock_structure}.
Due to efficient radiative cooling, this shocked region forms a geometrically thin, dense shell \citep[e.g.][]{Chugai2004-xx}, often termed the ``cold dense shell.''
Assuming that the thickness of the shocked shell is negligible compared to its radius, the dynamics of the shocked shell can be fully characterised by its mass $M_\sh (t)$ and radius $r_\sh (t)$.

The evolution of these parameters over time is governed by conservation laws. Mass conservation gives:
\begin{equation} \label{eq:mass_conservation}
    \odv{M_\sh}{t} = 4 \pi {r_\sh}^2 [\rho_\ej (v_\ej - v_\sh) + \rho_\CSM (v_\sh - v_\CSM)],
\end{equation}
and momentum conservation gives: 
\begin{equation} \label{eq:momentum_conservation}
    M_\sh \odv{v_\sh}{t} = 4 \pi {r_\sh}^2 [\rho_\ej (v_\ej - v_\sh)^2 - \rho_\CSM (v_\sh - v_\CSM)^2],
\end{equation}
where $v_\sh := \odif{r_\sh} / \odif{t}$ represents the velocity of the shocked shell, $\rho_\ej$ and $v_\ej$ denote the ejecta density and velocity near the shocked shell, $\rho_\CSM$ and $v_\CSM$ denote the CSM density and velocity near the shocked shell. We assume the density structure of the CSM to be in the form of a power law:
\begin{equation} \label{eq:CSM_density}
    \rho_\CSM = D r^{-s}.
\end{equation}

We further assume that the SN ejecta is in the homologous expansion phase and has a broken power-law density structure \citep[e.g.,][]{Chevalier1989-zd}, given by
\begin{equation} \label{eq:ejecta_density}
    \rho_\ej (v_\ej, t) = A 
    \begin{cases}
        t^{-3} (v / v_\tr)^{- \delta} & (v > v_\tr) \\
        t^{-3} (v / v_\tr)^{- n} & (v < v_\tr)
    \end{cases}
\end{equation}
and
\begin{equation}
    v_\tr = \sqrt{\frac{2 (5 - \delta) (n - 5) E_\ej}{(3 - \delta)(n - 3) M_\ej}}.
\end{equation}
Here, $E_\ej$ and $M_\ej$ represent the total kinetic energy and total mass of the SN ejecta, respectively.
The value of $A$ is determined such that the integral of $\rho_\ej$ gives $M_\ej$.
These assumptions are based on results from numerical simulations \citep[e.g.][]{Matzner1999-lt}, which also indicate that typically, $\delta = 0 - 1$ and $n = 7 - 12$, depending on the progenitor's nature.

Using the density structures of the CSM (Equation \ref{eq:CSM_density}) and the SN ejecta (Equation \ref{eq:ejecta_density}), we can solve Equation (\ref{eq:mass_conservation}) and Equation (\ref{eq:momentum_conservation}) for the evolution of the shocked shell.
If the CSM is interacting with the outer ejecta, we asymptotically have \citep{Moriya2023-kk},
\begin{equation} \label{eq:shell_rad}
    r_\sh (t) = \left[ \frac{(3-s) (4-s)}{4 \pi D (n-4) (n-3) (n-\delta)}
    \frac{[2 (5-\delta) (n-5) E_\ej]^{(n-3)/2}}{[(3-\delta) (n-3) M_\ej]^{(n-5)/2}} \right]^{1/n} t^{(n-3)/(n-s)}.
\end{equation}

A portion of kinetic energy dissipated by collision with the shocked shell is converted into radiation.
The total luminosity produced via this interaction can be expressed as
\begin{equation}
    \Lint = \varepsilon \odv{E_\mathrm{kin}}{t} = 2 \pi \varepsilon {r_\sh}^2 [\rho_\ej (v_\sh - v_\ej)^3 + \rho_\CSM (v_\sh - v_\CSM)^3] \label{eq:lum_int_general}
\end{equation}
where $\varepsilon$ is the conversion efficiency from kinetic to radiation energy, and $E_\mathrm{kin}$ is the total kinetic energy of the unshocked ejecta and the unshocked CSM.
We assume $\varepsilon = 0.3$ here \citep[e.g.][]{Fransson2014-tv}.
The first term, $L_\mathrm{RS} = 2 \pi \varepsilon {r_\sh}^2 \rho_\ej (v_\sh - v_\ej)^3$, represents the luminosity from the reverse shock, while the second term, $L_\mathrm{FS} = 2 \pi \varepsilon {r_\sh}^2 \rho_\CSM (v_\sh - v_\CSM)^3$, represents the luminosity from the forward shock.

Under the assumptions $v_\sh \gg v_\CSM$ and $L_\mathrm{FS} \gg L_\mathrm{RS}$, the luminosity evolution while the CSM is interacting with the outer ejecta is given by
\begin{multline} \label{eq:int_lum}
    \Lint (t) = \frac{\varepsilon}{2} (4 \pi D)^{(n-5)/(n-s)} \left(\frac{n-3}{n-s}\right)^3 \left[ \frac{(3-s) (4-s)}{(n-4) (n-3) (n-\delta)}\right]^{(5-s)/(n-s)} \\
    \left[ \frac{[2 (5-\delta) (n-5) E_\ej]^{(n-3)/2}}{[(3-\delta) (n-3) M_\ej]^{(n-5)/2}} \right]^{(5-s)/(n-s)} t^{(6s - 15 + 2n - ns) / (n-s)} .
\end{multline}

If $n = 10$, $\delta = 0$, and the CSM density distribution is flat ($s = 0$), Equation (\ref{eq:int_lum}) implies $\Lint (t) \propto t^{1/2}$.
Therefore, if the CSM is optically thin enough that the light curve is not altered by the effects of diffusion, the light curve keeps rising as long as the outer ejecta is interacting with the CSM.
This contrasts with the case of a wind-like density structure of CSM ($s = 2$), in which for $n = 10$ and $\delta = 0$, Equation (\ref{eq:int_lum}) gives $\Lint (t) \propto t^{-3/8}$ and decreases over time. 

From now on, we limit our scope to the ``flat'' configurations of the CSM; those with small enough $s$ such that the exponent of $t$ in the right hand side of Equation (\ref{eq:int_lum}) is positive.
If the CSM is massive enough (see Section \ref{sec:lc_classification}), the inner SN ejecta eventually reaches the shocked shell, after which the interaction luminosity begins to decline.
Here, we refer to the onset of the interaction between the inner ejecta and the CSM as the moment of ``transition''.
If transition occurs, the peak interaction luminosity is attained at the time of the transition, and it is given by
\begin{equation} \label{eq:trans_time}
    t_\tr = \left[ \frac{(3-s) (4-s)}{4 \pi D (n - 4)(n - 3)(n - \delta)} \frac{[3 - \delta)(n - 3) M_\ej]^{(5-s)/2}}{[2 (5 - \delta) (n - 5) E_\ej]^{(3-s)/2}} \right]^{1/(3-s)}
\end{equation}


However, transition does not occur if the CSM radius is not massive enough the shocked shell reaches the edge of the CSM before transition.
In this case, $\Lint (t)$ increases monotonically throughout the interaction with a flat CSM.
Following \citet{Khatami2023-oy}, here, we refer to this an ``emergence'' of the shock.
If the shock emerges from the CSM before transition, the interaction luminosity peaks at the time of emergence instead.
The time of emergence is obtained by solving $r_\sh(t) = R_\CSM$ for $t$ with Equation (\ref{eq:shell_rad}) as $r_\sh (t)$;
\begin{equation} \label{eq:emer_time}
    \tem = \left[ \frac{4 \pi D (n-4) (n-3) (n-\delta)}{(3-s) (4-s)}
    \frac{[(3-\delta) (n-3) M_\ej]^{(n-5)/2}}{[2 (5-\delta) (n-5) E_\ej]^{(n-3)/2}} \right]^{(n-s) / [n (n-3)]} {R_\CSM}^{(n-s) / (n-3)}.
\end{equation}

\newpage


\subsection{Effects of Diffusion} \label{sec:diffusion}

We now proceed to present our method for describing the effects of photon diffusion on the light curve, which was not included the model of \citet{Moriya2023-kk}. The opacity of the CSM is assumed to be constant, appropriate for Thomson scattering. We also assume $s = 0$ (and hence, $\rho_\CSM = D$, a constant) for simplification.

Diffusion of photons inside the CSM is characterised by the optical depth of the CSM.
The total radial optical depth of unshocked CSM from the shell to the edge of the CSM at time $t$ is given by
\begin{equation}
    \tau_\CSM (t) = \kappa_\CSM D [R_\CSM - r_\sh (t)]
\end{equation}
where $\kappa_\CSM$ is the opacity of CSM and $R_\CSM$ is the total radius of CSM.

If the CSM is optically thick enough, photons are overtaken by the shocked shell while it is diffusing outwards and therefore cannot escape the system.
Once the forward shock front reaches the radius where $\tau_\CSM$ is low enough, photons can diffuse ahead of the shocked shell, and the light curve starts rising.
This phenomenon is termed ``shock breakout''.
\footnote{It should be noted that, in some conventions, the term ``shock breakout'' refers to the arrival of the shock at the edge of the CSM. This definition is not adopted here.}
Ignoring a factor of order unity, the shock breakout occurs when the following condition is satisfied:
\begin{equation} \label{eq:breakout_cond}
    \tau_\CSM (t) = \frac{c}{v_\sh (t)}.
\end{equation}
In this paper, we use the time when Equation (\ref{eq:breakout_cond}) is satisfied, $t_\br$, to estimate the condition for the breakout to occur.

Equation (\ref{eq:breakout_cond}) can be satisfied if the total optical depth of the CSM, $\tau_\CSM (0)$, exceeds $c / v_\sh (0)$.
In this case, the rise time of the light curve is comparable to the timescale for photons to diffuse out of the CSM.
If the CSM is optically thick ($\tau_\CSM (t) \gg 1$), the diffusion timescale of the unshocked CSM at time $t$ can be expressed as
\begin{equation} \label{eq:diffusion_timescale}
    t_\diff (t) = \frac{\kappa_\CSM D [R_\CSM - r_\sh (t)]^2}{c}.
\end{equation}
Since the CSM is optically thick at the time of the shock breakout ($\tau_\CSM(t_\br) = c / v_\sh \sim 30$), the rise time can be approximated by $t_\diff (t_\br)$.

The luminosity due to the breakout can be estimated as follows.
Since photons produced by the interaction during $t = 0$ to $t = t_\br$ are trapped at the shock front, the total energy of radiation produced by the interaction that is released with the breakout is
\begin{equation}
    E_\br = \int_0^{t_\br} \Lint (t) \odif{t}.
\end{equation}
Since this amount of energy is emitted on the timescale of $t_\diff (t_\br)$, we estimate the luminosity resulting from shock breakout as
\begin{equation} \label{eq:breakout_lum}
    L_\br = \frac{1}{t_\diff(t_\br)} \int_0^{t_\br} \Lint (t) \odif{t}.
\end{equation}
Then, the luminosity of the SN at $t = t_\br + t_\diff(t_\br)$ when the contribution of the breakout to the SN luminosity is maximal is given by $L_\br + \Lint (t_\br)$. If $L_\br$ is comparable to or larger than luminosity from interaction, the breakout create a peak in the light curve.

If $\tau_\CSM (0) < c / v_\sh(0)$, the shock breakout does not occur while the shock is located in the CSM, and the rise time is comparable to the timescale of diffusion through the whole CSM,
\begin{equation}
    t_\diff (0) = \frac{\kappa_\CSM D {R_\CSM}^2}{c},
\end{equation}
instead. Here, we assumed that the radius of the progenitor is much smaller than that of the CSM.

\section{Morphology of Interaction-powered Light Curves from flatly distributed CSM} \label{sec:morphology}

\subsection{Classification of Light Curves} \label{sec:lc_classification}

\begin{figure}[h!]
    \centering
    \includegraphics[width = 0.90 \textwidth]{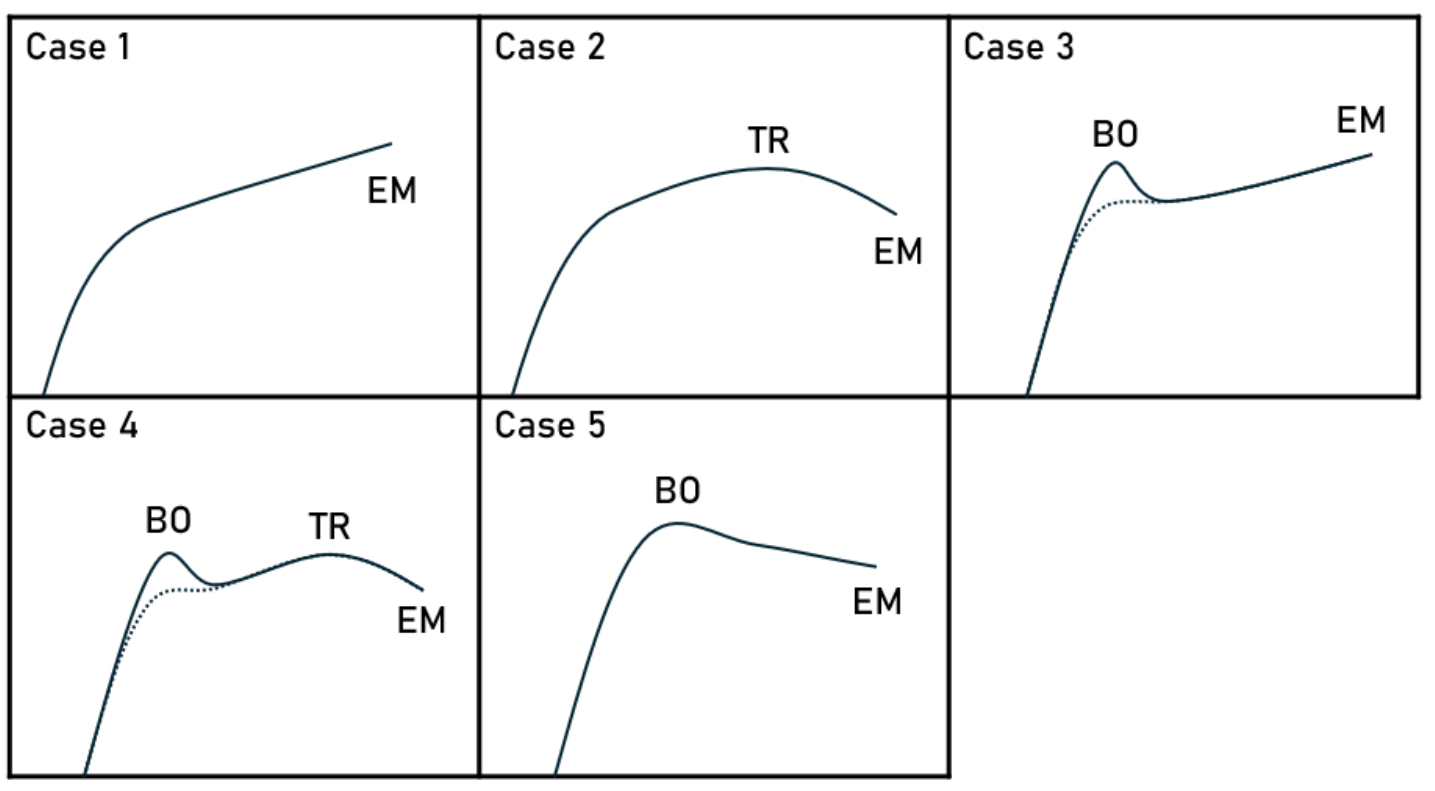}
    \caption{The qualitative shapes of light curves for each of the five classes.
    Captions represent the physical processes corresponding to the features of light curves (BO = shock breakout, TR = transition, EM = Emergence).
    Dotted lines show the shape of the light curve if the luminosity due to breakout is moderate and the light curve has only one peak.}
    \label{fig:lc_shapes}
\end{figure}

In our framework with flatly distributed CSM, three different mechanisms create peaks in light curves: shock breakout, transition, and emergence.
For the given explosion parameters, $E_\ej$ and $M_\ej$, the values of $M_\CSM$ and $R_\CSM$ determine which of those processes occur during the ejecta--CSM interaction.
The light curves can be categorised into following five classes, based on $t_\tr$ and $\tem$, the times of transition and shock emergence (both defined in Section \ref{sec:evolution}), and $t_\br$, the time of breakout (defined in Section \ref{sec:diffusion}).
Figure \ref{fig:lc_shapes} illustrates the general shapes of predicted light curves that fall under each of the classes.

\begin{itemize}
    \item \textbf{Case 1}: No transition ($\tem < t_\tr$), no breakout ($\tau_\CSM < \tau_\crit$)
    
    In this case, the CSM shell is optically too thin for the breakout to occur while the shock travels through the CSM.
    Photons produced by the ejecta--CSM interaction starts to escape the system right after the onset of the interaction.
    The light curve rises due to photons diffusing out of the CSM at first, and then to due to the steady increase of the interaction luminosity.

    \item \textbf{Case 2}: With transition ($\tem > t_\tr$), no breakout ($\tau_\CSM > \tau_\crit$)
    
    Similarly to the previous case, the light curve rises due to photons diffusing out of the CSM.
    However, in this case the interaction luminosity increases only while $t < t_\tr$.
    Since the photons produced in the shell at $t = t_\tr$ escape the CSM at, on average, $t = t_\tr + t_\diff(t_\tr)$, the light curve starts to decline around $t = t_\tr + t_\diff(t_\tr)$.

    \item \textbf{Case 3}: No transition ($\tem < t_\tr$), with breakout ($\tau_\CSM < \tau_\crit$)
    
    In this case, photons start diffusing out only after the shock breakout.
    If $L_\br$ is large enough, the breakout can create a peak in the light curve.
    Subsequently, interaction luminosity keeps rising until $t = \tem$.

    \item \textbf{Case 4}: With transition ($\tem > t_\tr$), with breakout ($\tau_\CSM < \tau_\crit$), $t_\br < t_\tr$
    
    As with the last case, the light curve starts to rise only after the shock breakout, and the light curve can exhibit a peak due to the breakout.
    Afterwards, interaction luminosity increases while $t < t_\tr$ and then decreases.
    If $L_\br$ is large enough, the light curve displays a peak at around $t = t_\tr + t_\diff(t_\tr)$.
    
    \item \textbf{Case 5}: With transition ($\tem > t_\tr$), with breakout ($\tau_\CSM < \tau_\crit$), $t_\tr < t_\br$
    
    In this case, after the shock breakout, the interaction luminosity decreases with time.
    Therefore, the light curve has its only peak due to the shock breakout.
\end{itemize}

Notably, in Case 3 and Case 4, light curves can exhibit two peaks; one due to the shock breakout and the other due to change in the interaction luminosity.
This possibility is investigated further in Section~\ref{sec:observation}.

\begin{figure}[h!]
    \centering
    \includegraphics[width = 0.75 \textwidth]{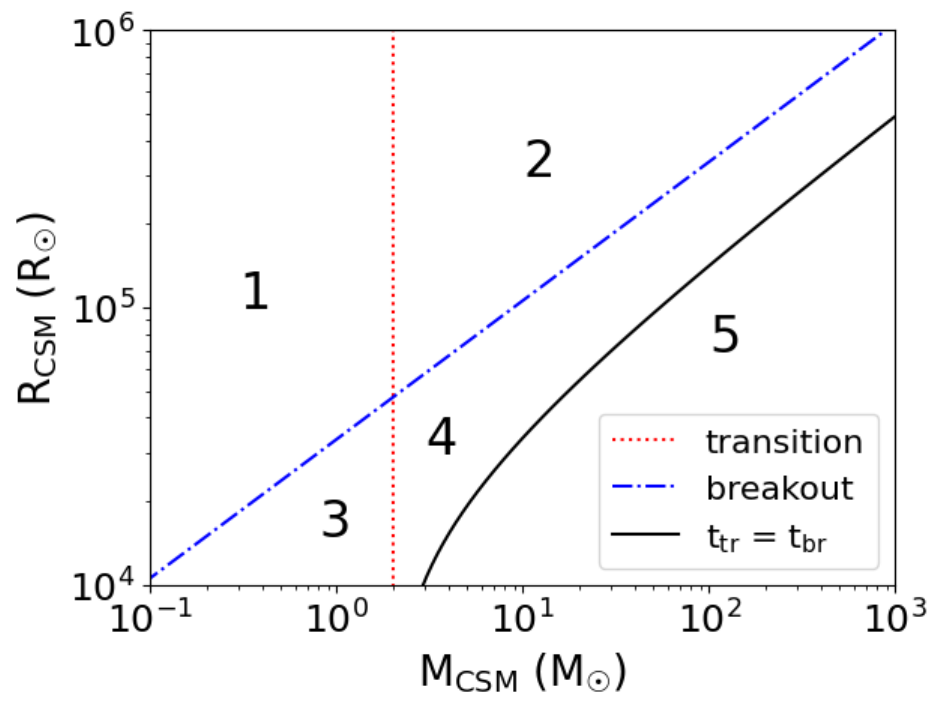}
    \caption{Boundaries between the five classes of light curves in the $(M_\CSM, R_\CSM)$ space.
    The numbers inside the figure correspond to the numbers given to the classes of light curves in Figure \ref{fig:lc_shapes}.
    The power indices of $n = 10$, $\delta = 0$, the explosion energy of $E_\ej = 10^{51} \ \mathrm{erg}$, the ejecta mass of $M_\ej = 10 \ \Msun$, and the CSM opacity of $\kappa_\CSM = 0.34 \ \mathrm{cm^2 \ g^{-1}}$ are assumed.}
    \label{fig:lc_cases}
\end{figure}

Figure \ref{fig:lc_cases} shows an example of the boundaries between each class in the $(M_\CSM, R_\CSM)$ space.
In this example, the density structure of the ejecta is assumed to have $n = 10$, $\delta = 0$, and the explosion energy and the mass of the ejecta are set to $E_\ej = 10^{51} \ \mathrm{erg}$ and $M_\ej = 10 \ \Msun$, respectively.
The CSM velocity is set to $v_\CSM = 100 \ \mathrm{km \ s^{-1}}$, and the opacity of the CSM is set to $\kappa_\CSM = 0.34 \ \mathrm{cm^2 \ g^{-1}}$, the opacity of fully ionised, hydrogen-rich matter.

The boundary lines in Figure \ref{fig:lc_cases} are defined as follows:

\begin{itemize}
    \item \textbf{The dotted line} represents the boundary between whether a transition occurs or not.
    Transition occurs before emergence if $r_\sh (t_\tr) < R_\CSM$.
    In the case of $s = 0$, from Equations (\ref{eq:shell_rad}) and (\ref{eq:trans_time}), this condition is equivalent to
    \begin{equation} \label{eq:with_trans_cond}
        M_\CSM > M_\crit := \frac{4 (3 - \delta)}{(n - 4) (n - \delta)} M_\ej.
    \end{equation}
    In the example, $M_\crit = M_\ej / 5 = 2 \ \Msun$.
    It is worth mentioning that $M_\crit$ is independent of $R_\CSM$ and $E_\ej$.

    \item \textbf{The dot-dash line} is the approximate location of the boundary between whether a breakout occurs or not.
    Shock breakout occurs when
    \begin{equation} \label{eq:with_breakout_cond}
        \tau_\CSM = \frac{3 \kappa_\CSM M_\CSM}{4 \pi {R_\CSM}^2} > \tau_\crit = c / v_\crit,
    \end{equation}
    for some $v_\mathrm{crit}$ comparable the initial shock velocity in the CSM, $v_\sh (0)$.
    Here, we assumed a fiducial value of $v_\crit = 10000 \ \mathrm{km \ s^{-1}}$, or $\tau_\crit = 30$.
    
    \item \textbf{The solid line} indicates the choices of $(M_\CSM, R_\CSM)$ such that $t_\br = t_\tr$.
    To the left (right) of the line, we have $t_\br < t_\tr$ ($t_\br > t_\tr$). 
\end{itemize}

\subsection{Characteristic Timescale and Luminosity of Light Curves} \label{sec:time_luminosity}

Based on the classification scheme in the previous section, we quantitatively investigate the properties of the light curves in our model.
We make the following assumptions as order-of-magnitude estimates:
\begin{itemize}
    \item The time of the shock breakout, $t_\br$, is given as the solution of Equation (\ref{eq:breakout_cond}).

    \item The duration of the rise in the light curve is given by $t_\diff (t_\br)$, the diffusion timescale of the unshocked CSM (Equation \ref{eq:diffusion_timescale}) at $t = t_\br$.

    \item The maximal luminosity due to the shock breakout is given by Equation (\ref{eq:breakout_lum}).

    \item The peak in the interaction luminosity at $t = t_\tr$ appears in the light curve at
    \begin{equation}
    t = 
        \begin{cases}
            t = t_\tr + t_\diff (t_\tr) & (\tau_\CSM(t_\tr) > 1), \\
            t = t_\tr + [R_\CSM - r_\sh (t_\tr)] / c & (\tau_\CSM(t_\tr) < 1).
        \end{cases}
    \end{equation}
\end{itemize}

\begin{figure}[h!]
    \centering
    \includegraphics[width = 0.9 \textwidth]{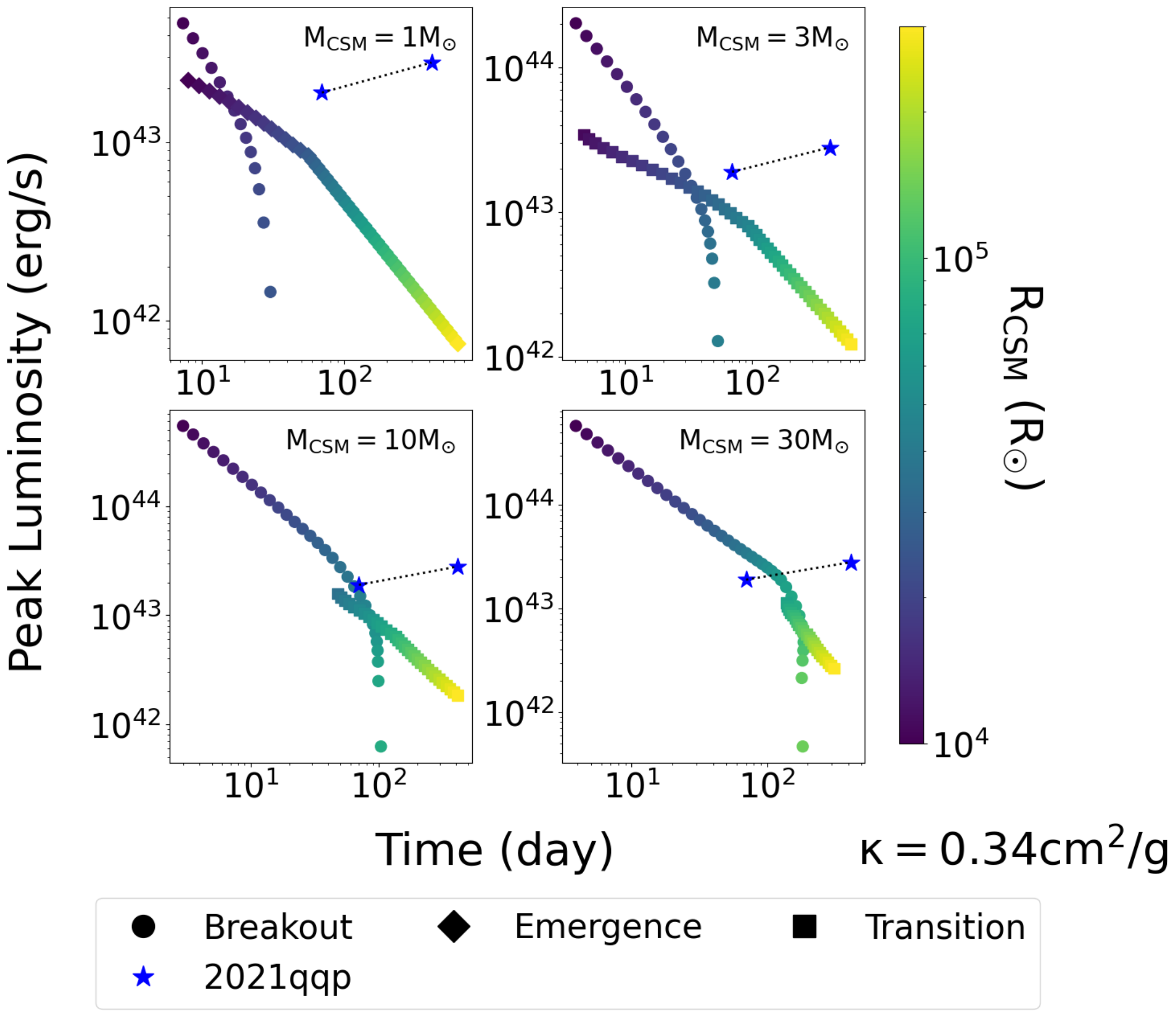}
    \caption{Timescales and luminosities of light curve features for various CSM masses.
    The names of light curve features are defined in Section \ref{sec:method} and illustrated in Figure \ref{fig:lc_shapes}.
    The power indices of $n = 10$, $\delta = 0$, the CSM opacity of $\kappa_\CSM = 0.34 \ \mathrm{cm^2 \ g^{-1}}$, the explosion energy of $E_\ej = 10^{51} \ \mathrm{erg}$, and the ejecta mass of $M_\ej = 10 \ \Msun$ (and hence $M_\crit = 2 \ \Msun$) are assumed. 
    For comparison, approximate times and luminosities of two peaks in the light curve of Type IIn SN 2021qqp \citep{Hiramatsu2024-hb} are shown.}
    \label{fig:time_luminosity_mcsm}
\end{figure}

Figure \ref{fig:time_luminosity_mcsm} shows how the timescales and luminosities of the light curve features change with respect to CSM radius, $R_\CSM$.
Time is measured from the ``first light,'' or the moment when the light curve begins to rise; $t = 0$ (without breakout), or $t = t_\br$ (with breakout).
In this example, the power indices of the ejecta are assumed to have the values $n = 10$, $\delta = 0$, the explosion energy $E_\ej = 10^{51} \ \mathrm{erg}$, and the mass $M_\ej = 10 \ \Msun$.
In this setting, as seen in Equation (\ref{eq:with_trans_cond}), transition occurs before emergence if $M_\CSM < 2 \ \Msun$.
Furthermore, the opacity of the CSM is set to $\kappa_\CSM = 0.34 \ \mathrm{cm^2 \ g^{-1}}$.
Four values of CSM mass are considered: $M_\CSM = 1 \ \Msun$ (with transition) and $M_\CSM = 3, 10, 30 \ \Msun$ (without transition).
In all of these cases, rise times longer than 100 days are achievable when the CSM extends for more than $\sim 10^5 \ \Rsun$.
We can also see that for a fixed CSM radius, if the CSM is more massive, the light curve is brighter and evolves faster.

As noted in the previous section, for each value of CSM mass, there exists a range of CSM radii that results in double-peaked light curves.
For example, in the case of $M_\CSM = 10 \ \Msun$, such a light curve arises if the radius of the CSM is $\sim 5 \times 10^4 \ \Rsun$, and the light curve evolves in the timescale of $\sim 100$ days.

\begin{figure}[h!]
    \centering
    \includegraphics[width = 0.9 \textwidth]{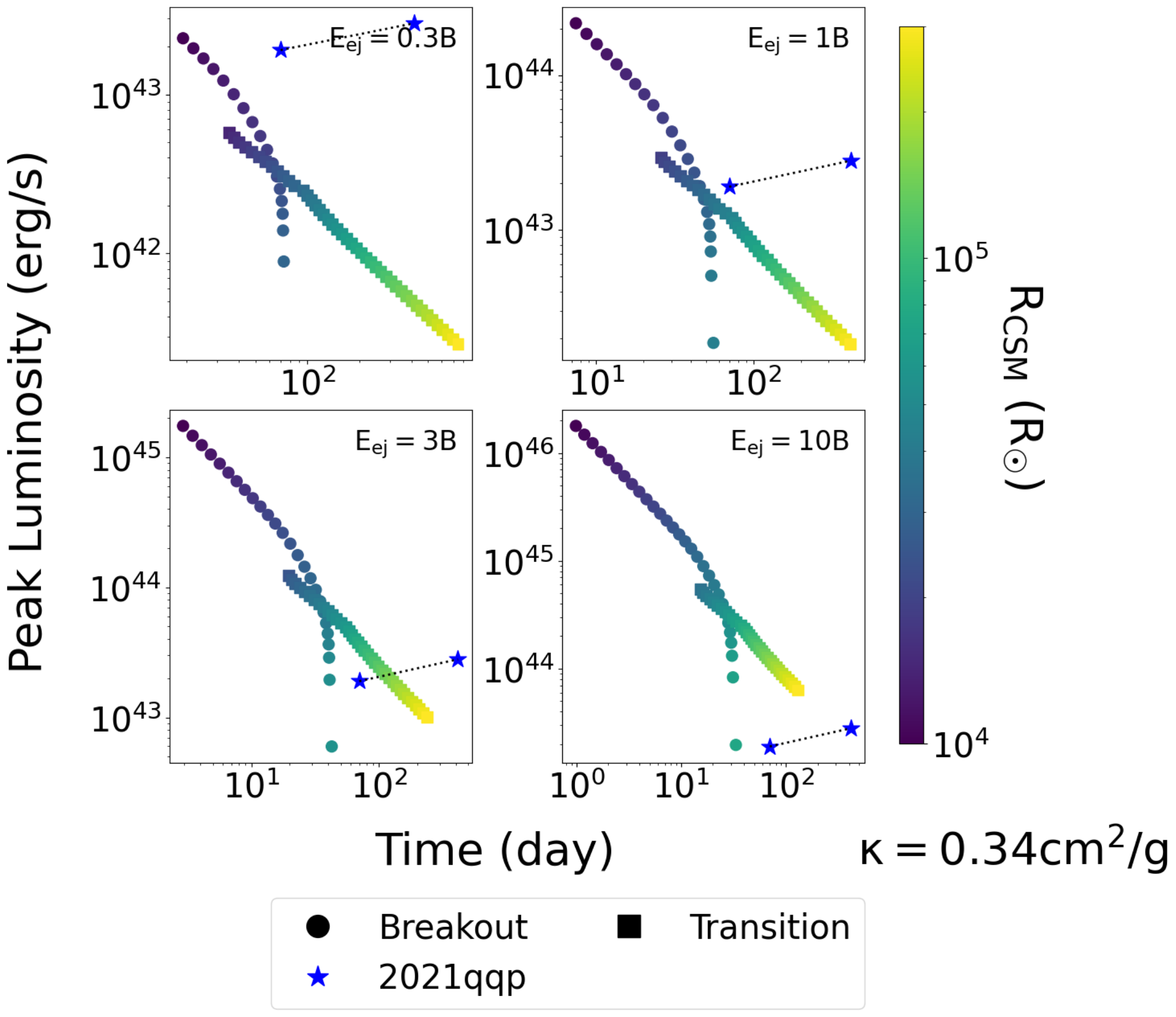}
    \caption{Same as Figure \ref{fig:time_luminosity_mcsm}, but for various ejecta energies.
    The power indices of $n = 10$, $\delta = 0$, the CSM opacity of $\kappa_\CSM = 0.34 \ \mathrm{cm^2 \ g^{-1}}$, the ejecta mass of $M_\ej = 10 \ \Msun$, and the CSM mass of $M_\CSM = 10 \ \Msun$ are assumed.}
    \label{fig:time_luminosity_eej}
\end{figure}

\begin{figure}[h!]
    \centering
    \includegraphics[width = 0.9 \textwidth]{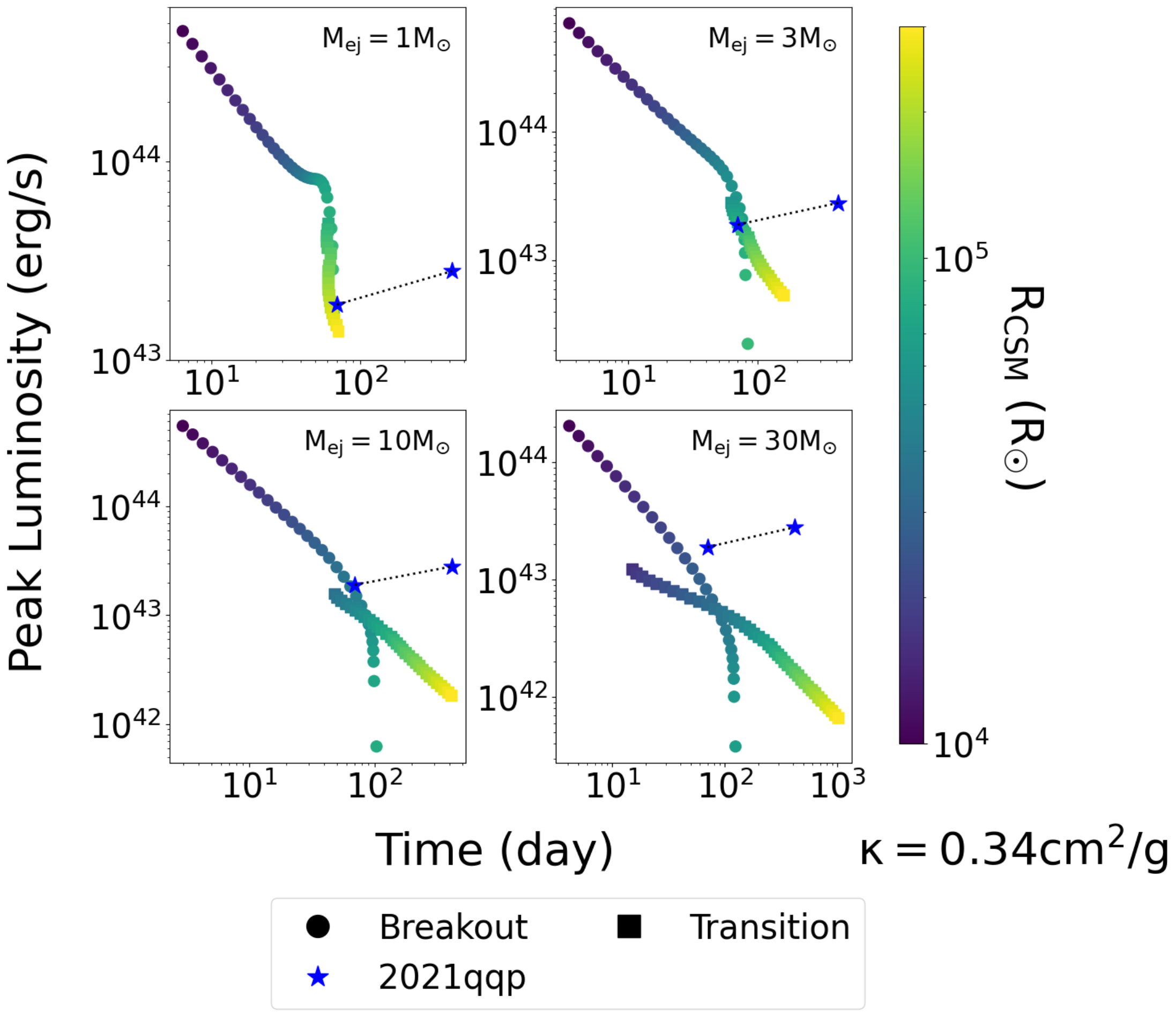}
    \caption{Same as Figure \ref{fig:time_luminosity_mcsm}, but for various ejecta masses.
    The power indices of $n = 10$, $\delta = 0$, the CSM opacity of $\kappa_\CSM = 0.34 \ \mathrm{cm^2 \ g^{-1}}$, the explosion energy of $M_\ej = 10^{51} \ \mathrm{erg}$, and the CSM mass of $M_\CSM = 10 \ \Msun$ are assumed.}
    \label{fig:time_luminosity_mej}
\end{figure}

Figures \ref{fig:time_luminosity_eej} and \ref{fig:time_luminosity_mej} illustrate how the timescales and luminosities of the peaks depend on the explosion energy, $E_\ej$, and the ejecta mass, $M_\ej$, respectively.
As seen in Figure \ref{fig:time_luminosity_eej}, a more energetic explosion produces brighter light curves with slightly faster timescales.
Figure \ref{fig:time_luminosity_mej} demonstrates that more massive ejecta results in a similar peak luminosity but with slower evolution, particularly for the second peak due to transition or emergence.
The dependence of evolution timescales to the parameters $E_\ej$ and $M_\ej$ can also be seen from Equations (\ref{eq:trans_time}) and (\ref{eq:emer_time}); the time of transition, $t_\tr$, and the time of emergence, $\tem$, depend on $E_\ej$ and $M_\ej$ as $t_\tr \propto {M_\ej}^{5/6} {E_\ej}^{-1/2}$ and $\tem \propto {M_\ej}^{5/14} {E_\ej}^{-1/2}$, with the choices of power indices adopted here, $\delta = 0$, $n = 10$, and $s = 0$.

\begin{figure}[h!]
    \centering
    \includegraphics[width = 0.9 \textwidth]{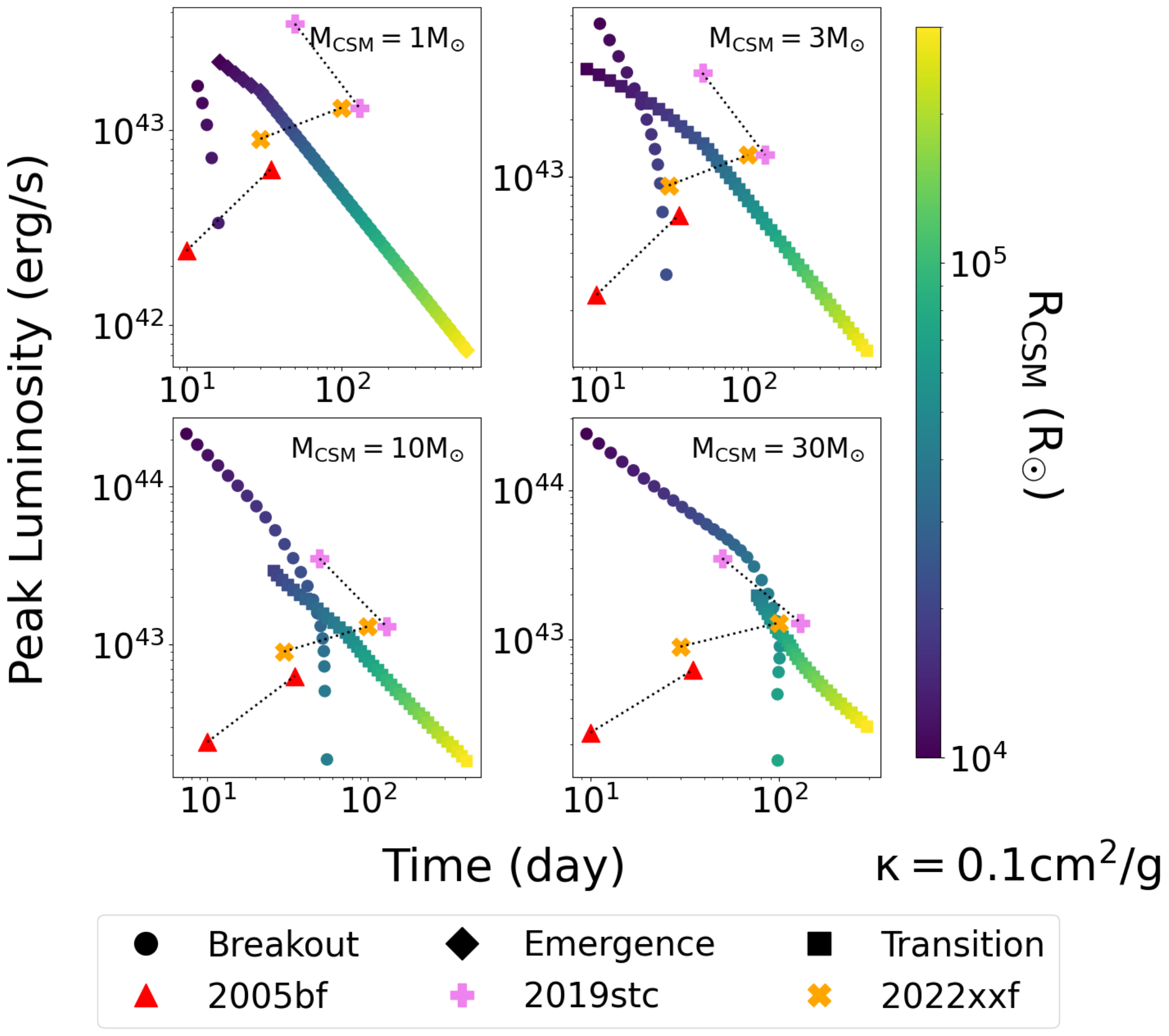}
    \caption{Same as Figure \ref{fig:time_luminosity_mcsm}, but using a different value of opacity, $\kappa_\CSM = 0.1 \ \mathrm{cm^2 \ g^{-1}}$. \\
    For comparison, approximate times and luminosities of two peaks in the light curves of three stripped-envelope SNe, SN 2005bf \citep{Folatelli2006-ar}, SN 2019stc \citep{Gomez2021-ev}, and SN 2022xxf \citep{Kuncarayakti2023-vd}, are shown.}
    \label{fig:time_luminosity_kappa}
\end{figure}

Figure \ref{fig:time_luminosity_kappa} is the same as Figure \ref{fig:time_luminosity_mcsm}, but the values are obtained using a different value for the CSM opacity, $\kappa_\CSM = 0.1 \ \mathrm{cm^2 \ g^{-1}}$.
This value corresponds to hydrogen-poor circumstellar environments typical of progenitor stars for SNe of Type Ibn or Icn.
When comparing Figure \ref{fig:time_luminosity_kappa} to Figure \ref{fig:time_luminosity_mcsm}, the main difference is the peak luminosity due to the breakout, which decreases with a lower value of $\kappa_\CSM$.
This relationship can be explained by Equation (\ref{eq:breakout_cond}), the condition for breakout.
The optical depth of the unshocked CSM, $\tau_\CSM$, is proportional to the opacity of the CSM, $\kappa_\CSM$.
Since changing $\kappa_\CSM$ does not affect the motion of the shocked shell through the CSM, reducing $\kappa_\CSM$ causes the breakout to occur earlier.
This means that the amount of energy trapped at the shocked shell before the breakout is diminished, resulting in a lower luminosity of the peak due to the breakout.

\subsection{Comparison with Known Double-peaked Light Curves} \label{sec:observation}

We further investigate the possibility of slowly evolving, double-peaked light curves within our model with a flatly distributed CSM.
We also compare our theoretical results with observations of such double-peaked SNe from the literature.

First, it should be noted that for a doubly peaked light curve with comparable peak luminosities to be well reproduced in our model, the duration between the two peaks has to be shorter than or comparable to the rise time of the first peak.
This can be shown qualitatively using Equation (\ref{eq:breakout_lum}).
Since $\Lint$ increases with time, for the right-hand side of Equation (\ref{eq:breakout_lum}) to be comparable to the interaction luminosity at the time of transition, $t = t_\tr$, the time of breakout, $t_\br$, must be at least comparable to $t_\tr$.
If the two peaks are too temporally separated, the breakout luminosity $L_\br$ cannot be comparable to the luminosity of the second peak.
This is also the reason why this model struggles to explain the behaviour of a light curve if its first peak is significantly brighter than its second (as in the case of SN 2019stc in Figure~\ref{fig:time_luminosity_kappa}).

From Figures \ref{fig:time_luminosity_mcsm} and \ref{fig:time_luminosity_eej}, we can see that, if a double-peaked light curve were to be realised in this model, the luminosity of the peaks would roughly correspond to the explosion energy, and the evolution timescale to the CSM and ejecta masses.
The evolution timescale of double-peaked light curves that can be explained by this model is at most $\lesssim 100$ days; to account for a long duration between peaks of a few hundred days (as in the case of SN 2021qqp in Figure~\ref{fig:time_luminosity_mcsm}), we would need to invoke unrealistically large ejecta mass of $\sim 1000 \ \Msun$.

Type~Ib SN 2005bf \citep{Folatelli2006-ar}, the prototype of a few other double-peaked SNe \citep{Taddia2018-fx, Gutierrez2021-yg}, and Type~Ic SN 2022xxf \citep{Kuncarayakti2023-vd}, whose two peaks are separated by 75 d, are examples of SNe whose light curves can be explained within our model.
Furthermore, the spectrum during the second rise of SN 2022xxf features narrow emission lines indicating the presence of CSM interaction.
\citet{Kuncarayakti2023-vd} attribute the first peak of SN 2022xxf to heating from $\hspace{0pt}^{56}\mathrm{Ni}$ decay and interpret the second as a result of the interaction between the SN ejecta and a detached CSM shell.
\citet{Takei2024-xu} analysed this interpretation in detail and found that the bolometric light curve of SN 2022xxf can indeed be well fit, but also pointed out that the CSM density in their model may be too low for the high-energy photons emitted from the shocked region to be converted to optical.
Here, we compare the light curves of these SNe with the predictions of our model, assuming that the interaction with flatly distributed CSM is responsible for both of the peaks in each of the light curves.
\footnote{The absence of interaction features can be reconciled with our model by considering, e.g., a clumpy distribution of the CSM.}

We assume power indices of the ejecta density profile to be $n = 10$, $\delta = 0$.
Since the SNe are of Type Ibc, we set the opacity of the CSM to be $\kappa_\CSM = 0.1 \ \mathrm{cm^2 \ g^{-1}}$.
For the CSM velocity, we consider $v_\CSM = 1000 \ \mathrm{km \ s^{-1}}$, the typical speed of the wind from Wolf-Rayet stars, which are commonly considered to be the progenitor of Type Ibc SNe.
By numerically solving the model presented in Section \ref{sec:method}, we sought the combination of parameters, SN properties ($E_\ej$, $M_\ej$) and CSM properties ($M_\CSM$, $R_\CSM$), that can reproduce the characteristics of the bolometric light curves, timescales and peak luminosities, of SN 2005bf and SN 2022xxf.
We have to bear in mind that our analyses here are based on order-of-magnitude estimates.
For example, when determining the time of the shock breakout, we used Equation (\ref{eq:breakout_cond}) and ignored the factor of order unity.
To gauge the uncertainty introduced by this assumption, we also looked for the combination of parameters that reproduce the light curves while assuming the factor of 2 in the breakout condition instead; that is, we redefined the moment of breakout as the moment when the equation $2 \tau_\CSM = c / v_\sh$ is satisfied.

The parameters we found are listed in Table \ref{tab:fit_param}.
Aside from the CSM radius, the parameters obtained for different breakout conditions vary significantly, although they are all of the same order of magnitude.
Table \ref{tab:obs_fit} compares the characteristic timescales and luminosities of the theoretical light curve using the parameters in Table \ref{tab:fit_param}.
We can see that each set of the estimated parameters can generally reproduce the timescales and luminosities of the observations.
The time of the first peak was not used in the estimation process, since it is difficult to tell the exact moment of explosion from light curves.
Nonetheless, Tables \ref{tab:fit_param} and \ref{tab:obs_fit} demonstrate the capability of our model to reproduce the characteristics of known SNe with double-peaked light curves.

The explosion energy and the ejecta mass inferred for SN 2022xxf in Table \ref{tab:fit_param} are much larger than those of typical SNe.
Pulsational pair-instability SN \citep[PPISN; e.g.][]{Woosley2007-ti, Yoshida2016-mw, Woosley2017-bo} is one possibility that can account for some of these values.
Violent pulses that occur $\sim 1 \ \mathrm{yr}$ before the explosion of some PPISN progenitors are expected to accompany significant mass loss of $\sim 10 \ \Msun$ \citep[see Figure 17 of][]{Woosley2017-bo}.
Additionally, pair-instability SN \citep[PISN; e.g.][]{Barkat1967-vu, Woosley2002-ok} can also give rise to the values similar to those inferred.
In the case of PISN, enhanced mass loss may be triggered by the progenitor's rapid rotation or the interaction with the progenitor's binary companion.

\begin{deluxetable}{cc|ccccc}[ht!]
    \tablecaption{Estimated parameters of the explosion for SNe 2005bf and 2022xxf.} \label{tab:fit_param}
    \tablehead{
        \colhead{} & 
        \colhead{Breakout} & 
        \colhead{$E_\ej$} & 
        \colhead{$M_\ej$} & 
        \colhead{$M_\CSM$} & 
        \colhead{$R_\CSM$} & 
        \colhead{$\tau_\CSM$} \\
        \colhead{} & 
        \colhead{condition} & 
        \colhead{($10^{51} \ \mathrm{erg}$)} & 
        \colhead{($\Msun$)} & 
        \colhead{($\Msun$)} & 
        \colhead{($\Rsun$)} &
        \colhead{}
    }
    \startdata
    \multirow{2}{*}{2005bf} & $\tau_\CSM = c / v_\sh$ & 1.05 & 14.0 & 3.20 & $2.2 \times 10^4$ & 65 \\
    & $2 \tau_\CSM = c / v_\sh$ & 0.58 & 4.75 & 1.15 & $ 2.0 \times 10^4$ & 28
    \\ \hline
    \multirow{2}{*}{2022xxf} & $\tau_\CSM = c / v_\sh$ & 9.1 & 165 & 37.9 & $6.4 \times 10^4$ & 94 \\
    & $2 \tau_\CSM = c / v_\sh$ & 4.7 & 52.4 & 16.7 & $6.3 \times 10^4$ & 40
    \enddata
\end{deluxetable}

\begin{deluxetable}{cc|cccc}[ht!]
    \tablecaption{Characteristic timescales and luminosities of the light curves corresponding to the parameters in Table \ref{tab:fit_param}.} \label{tab:obs_fit}
    \tablehead{
        \colhead{} & 
        \colhead{Breakout condition} &
        \colhead{First peak} & 
        \colhead{Between} & 
        \colhead{First peak} & 
        \colhead{Second peak} \\
        \colhead{} & 
        \colhead{} &
        \colhead{time (d)} & 
        \colhead{peaks (d)} & 
        \colhead{$L$ ($\mathrm{erg \ s^{-1}}$)} & 
        \colhead{$L$ ($\mathrm{erg \ s^{-1}}$)}
    }
    \startdata
    \multirow{2}{*}{2005bf} & $\tau_\CSM = c / v_\sh$ & 30 & 24 & $2.3 \times 10^{43}$ & $6.3\times 10^{43}$\\
     & $\tau_\CSM = 2 c / v_\sh$ & 12 & 25 & $2.4 \times 10^{43}$ & $6.3 \times 10^{43}$\\ \hline
    \multirow{2}{*}{2022xxf} & $\tau_\CSM = c / v_\sh$ & 109 & 76 & $0.8 \times 10^{43}$ & $1.4 \times 10^{43}$\\
    & $2 \tau_\CSM = c / v_\sh$ & 45 & 74 & $0.9 \times 10^{43}$ & $1.3 \times 10^{43}$\\
    \enddata
\end{deluxetable}

\section{Summary and Discussion} \label{sec:summary}

The interaction between SN ejecta and a flatly distributed CSM extends the diversity of the SN light curves in terms of timescales and morphology.
We have shown that the qualitative behaviour of light curves from SNe interacting with a CSM with a flat density profile can be characterised by the presence (or absence) and relative timing of three physical processes: breakout, transition and emergence.
Additionally, we have demonstrated that some configurations of the CSM shell lead to light curves with two peaks of similar luminosities over timescales of several months.
Furthermore, we showed that the timescales and luminosities of double-peaked SNe like SN 2005bf and SN 2022xxf can be well reproduced under this model with suitable SN and CSM parameters.

In this study, we assumed that SN light curves are dominated solely by interaction and ignored heating due to radioactive nickel.
The radioactive decay of nickel occurs on the timescale of a few tens of days.
In our model, it can contribute another peak to the light curve or make the peak due to the breakout brighter.
The latter implication is worth noting, since it can counter the difficulty in reproducing double-peaked light curves with its first peak brighter than the second in our model (Section \ref{sec:observation}).

We also focused solely on elucidating the qualitative behaviours of the bolometric light curves.
Future work should involve constructing detailed bolometric light curves and spectra by solving the radiative transfer equation, which will likely yield valuable insights into interacting SNe and stellar mass-loss events leading up to them.

Observations of double-peaked SN light curves are still scarce.
With the forthcoming advent of new facilities such as the Vera C. Rubin Observatory with enhanced capabilities to detect transients, we expect the diversity of SN light curves to broaden further.
Our results show the importance of CSM density profiles as a facet of CSM properties that gives rise to diverse SN light curves.
If new double-peaked light curves are observed, our model can be used as a simple method to constrain the CSM density profile.

\begin{acknowledgments}
We thank Nozomu Tominaga, Koh Takahashi, and Hanindyo Kuncarayakti for discussions. This work is supported by the Grants-in-Aid for Scientific Research of the Japan Society for the Promotion of Science (JP24K00682, JP24H01824, JP21H04997, JP24H00002, JP24H00027, JP24K00668) and by the Australian Research Council (ARC) through the ARC's Discovery Projects funding scheme (project DP240101786).
\end{acknowledgments}

\bibliography{paperpile}{}
\bibliographystyle{aasjournal}

\end{document}